\documentclass[12pt,preprint,showpacs]{revtex4}
\DeclareRobustCommand{\baselinestretch{2.2}}

\newcommand{\UFF}{Instituto de F\'{\i}sica da Universidade Federal Fluminense,\\ 
Boa Viagem 24210-340, Niter\'{o}i-RJ, Brazil.} 
\usepackage[]{graphicx}
\usepackage{bm}
\usepackage{amsmath}

\begin{document}

\title{Optical parametric oscillation under injection of orbital angular momentum}
\author{B. Coutinho dos Santos}
\email{bernardo@if.uff.br}
\author{C. E. R. Souza}
\email{ceduardo@if.uff.br}
\author{K. Dechoum}
\email{kaled@if.uff.br}
\author{A. Z. Khoury}
\email{khoury@if.uff.br}
\affiliation{\UFF}
\date{\today} 
\begin{abstract}
We present a theoretical model for the spatial mode dynamics of an optical 
parametric oscillator under injection of orbital angular momentum. This process 
is then interpreted in terms of an interesting picture based on a Poincar\'e 
representation of first order spatial modes. The spatial properties of the 
down-converted fields can be easily understood from their symmetries 
in this geometric representation. By considering the adiabatic mode conversion 
of the injected signal, we also predict the occurrence of a geometric phase 
conjugation in the down-converted beams. An experimental setup to measure this 
effect is proposed. 
\end{abstract}

\maketitle

\section{Introduction}
\label{intro}

Orbital angular momentum exchange in parametric amplification and oscillation has attracted a significant attention recently, both for its fundamental aspects as well as for its potential applications to quantum information processing. The first attempt to observe the transfer of orbital angular momentum (OAM) from the pump beam to the down converted beams was made by Arlt {\it et al.} in the spontaneous process, where the far field did not show the typical profile of an optical vortex.\cite{arlt}  Later, this problem was elucidated by Mair {\it et al} who investigated the same process in the photocount regime.\cite{Mair} They were able to demonstrate that the coincidence counts between twin photons generated by spontaneous parametric down conversion was subjected to OAM conservation. A further insight is given in ref. \onlinecite{Caetano}, where it was shown that parametric amplification in the stimulated process is also subjected to OAM conservation. There, an interpretation in terms of  transverse phase conjugation was provided.\cite{nossoprl} In ref.~\onlinecite{Martinelli}, the macroscopic transfer of OAM from the pump beam to one of the down converted beams was demonstrated in the operation of an optical parametric oscillator (OPO) above threshold. In that case, cavity mode selection was shown to play a crucial role in the OAM exchange.  

An important feature of optical beams bearing OAM is the geometric phase, or Berry's phase,\cite{berry} associated with cyclic transformations implemented with astigmatic mode converters,\cite{HELO,russos} first predicted by van Enk,\cite{Enk} and then experimentaly demonstrated by Galvez \textit{et al}.\cite{Galvez} This geometric phase is related to closed paths in a Poincar\'e sphere representation of optical beams,\cite{poincaresphere} and is the orbital equivalent of Pancharatnam's phase which is associated with cyclic transformations on the polarization state of light.\cite{pancharatnam}
There has been a renewed interest on geometric phases lately due to recent proposals of geometric quantum computation.\cite{quantcomp1,quantcomp2} 
The role of Pancharatnam's phase in nonclassical two-photon interference and the interplay between geometric and dynamical phase was discussed by J. Brendel \textit{et al}.\cite{brendel} They showed that the geometric phase acquired by the photon pair depends on the initial polarization state of the two photons.

The generation of discrete multidimensional entanglement by parametric down conversion with optical beams bearing OAM has been discussed in the literature.\cite{vaziri,steve,Barbosa}
In contrast to cavity free parametric down conversion, OPOs are an efficient source of continuous variable entanglement in the quadratures of intense optical beams.\cite{silvania,bowen} 
Therefore, parametric down conversion is a reliable source of quantum entanglement, while geometric phases are a potential tool for quantum computation. 
However, the properties of the geometric phase associated with OAM in parametric down-conversion has not been investigated yet.
In this article we start to combine these ideas by studying the classical properties of down converted beams generated by an injected OPO when adiabatic mode conversions are performed in the seed beam. An effect of geometric phase conjugation is predicted which is directly related to a symmetry between the down converted beams in the Poincar\'e representation.\cite{poincaresphere} We also propose an experimental setup for measuring this effect from the mutual interference between the down-converted beams. 
This geometric phase conjugation is equivalent to the time reversal of the Pancharatnam's phase as observed in four wave mixing by Tompkin \textit{et al}.\cite{chiao} Our work provides a first approach to the problem of geometric phase conjugation in parametric oscillation and opens the possibility for a future investigation in the quantum domain.

\section{Poincar\'e representation}
\label{Poincare}

\begin{figure}
\centering
\includegraphics[scale=0.7]{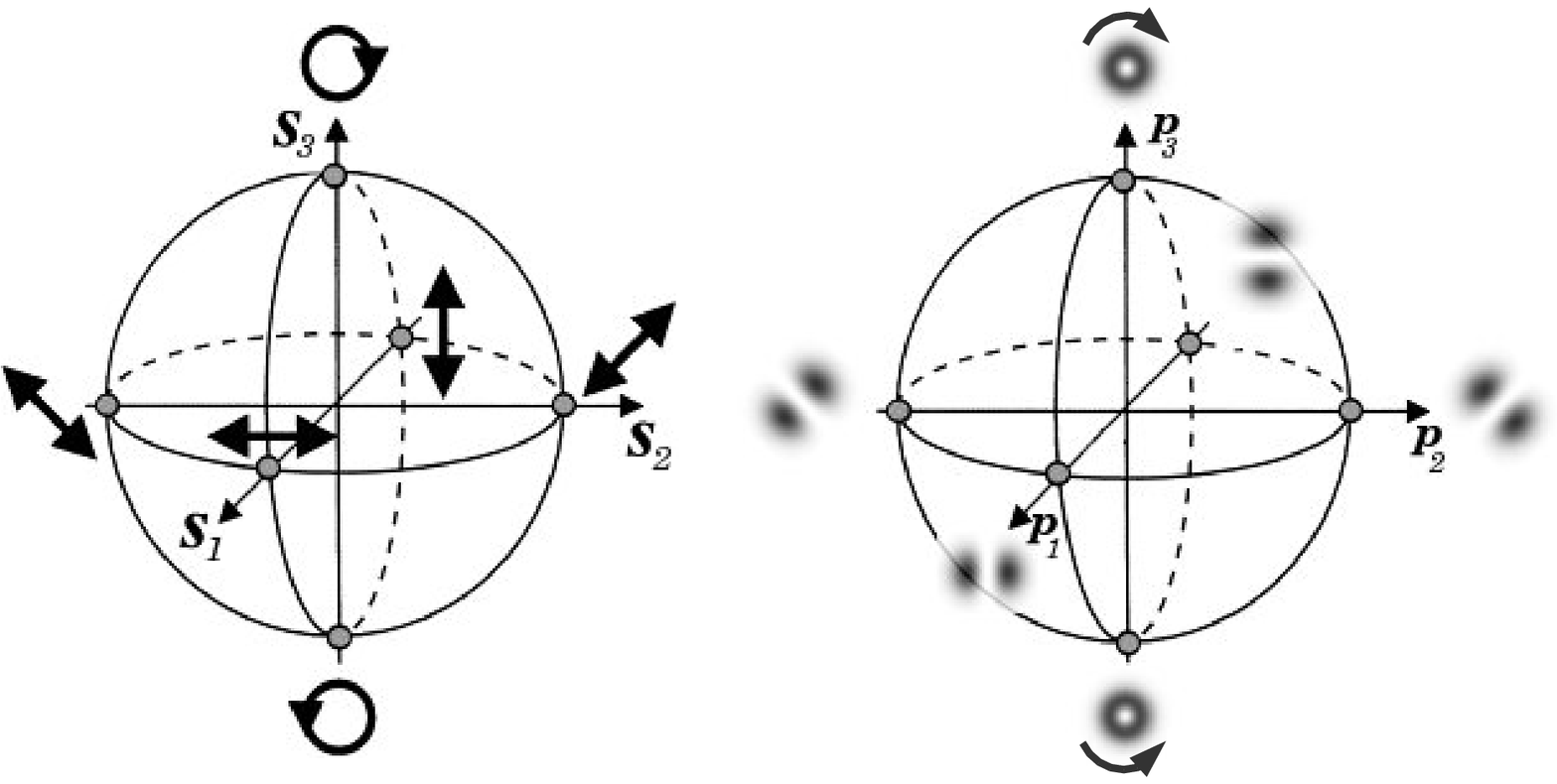}
\caption{Poincar\'e sphere for polarization states (left) and first order transverse modes (right). Notice the correspondence between circular polarization states and LG modes, placed on the poles, and linear polarization states and HG modes, placed on the equator.}
\label{fig:1}
\end{figure}

It is well known that polarization states of a monochromatic light beam can be completely caracterized by the Stokes parameters and mapped on the Poincar\'e sphere. As van Enk suggests in his paper,\cite{Enk} there is a correspondence between the polarization states and the first order transverse modes of the electromagnetic field. Based on this correspondence Padgett and Courtial propose a set of Stokes  parameters and an equivalent Poincar\'e representation for the first order transverse modes.\cite{poincaresphere}
\begin{eqnarray}\label{stoke1}
\label{p1}p_{1} &=& \dfrac{I_{HG_{0^{\circ}}}-I_{HG_{90^{\circ}}}}{I_{HG_{0^{\circ}}}+I_{HG_{90^{\circ}}}}\\
\label{p2}p_{2} &=& \dfrac{I_{HG_{45^{\circ}}}-I_{HG_{135^{\circ}}}}{I_{HG_{45^{\circ}}}+I_{HG_{135^{\circ}}}} \\
\label{p3}p_{3} &=& \dfrac{I_{LG_{+}}-I_{LG_{-}}}{I_{LG_{+}}+I_{LG_{-}}} \quad ,
\end{eqnarray} 
are the Stokes parameters describing a spatial profile belonging to the 
subspace of first order transverse modes. In a mode decomposition of the 
beam profile, $I_{\beta}$ is the square modulus of the coefficient 
of mode $\beta$. The first order modes are labeled so that 
$LG_{\pm}$ represents the first order Laguerre-Gaussian (LG) mode 
with topological charge $\pm 1$, respectively. $HG_{\theta}$ 
represents a first order Hermite-Gaussian (HG) mode rotated by an angle 
$\theta$ around the propagation axis. 
In fig. (\ref{fig:1}) we can see the Poincar\'e sphere for polarization 
states of a monochromatic beam and the equivalent representation for the 
first order transverse modes. The cartesian coordinates on the sphere are 
the respective Stokes parameters.

\section{Transverse mode dynamics}
\label{TMdyn}

\subsection{Equations of motion}
\label{eqmotion}

We now derive the dynamical equations for an optical parametric 
oscillator (OPO) operating under cavity conditions such that 
the pump laser is matched to a $TEM_{00}$ cavity mode and the 
down-converted beams are amplified with spatial profiles 
belonging to the first order modes subspace. This situation can 
be realized by a suitable tunning of the OPO cavity, combined with 
temperature control of the nonlinear crystal. Therefore, the 
intracavity pump field $E_{p}(\vec{r})$ can be written as 
\begin{equation}
E_{p}(\vec{r}) = \alpha_{p} \psi_{00}(\vec{\rho},z)\;,
\end{equation}
where $\alpha_{p}$ is the pump amplitude and 
$\psi_{00}(\vec{\rho},z)$ is the $TEM_{00}$ mode profile.

Now, let us decompose the intracavity signal and idler fields $E_{j}(\vec{r})$ 
($j=s,i$) into a linear superposition of first order LG 
modes: 
\begin{equation}
E_{j}(\vec{r}) = \alpha^{j}_{+} \psi_{+}(\vec{\rho},z) 
+ \alpha^{j}_{-} \psi_{-}(\vec{\rho},z) \quad ,
\end{equation} 
where $\psi_{\pm}(\vec{\rho},z)$ represent the spatial profiles 
of the first order LG modes with topological charges 
$\pm 1$ respectively, and $\alpha^{j}_{\pm}$ are the 
corresponding intracavity mode amplitudes.

We will consider the dynamics of an injected OPO where two 
input fields are sent to the OPO cavity. The pump input is 
mode matched to the $TEM_{00}$ mode and is described by the 
complex amplitude $\alpha^{p}_{in}\,$.
We also assume an incoming seed at the signal field which 
is prepared in a spatial mode corresponding to a given point 
in the Poincar\'e sphere. The input seed is then decribed 
by two complex amplitudes $\alpha_{+}^{s(in)}$ and 
$\alpha_{-}^{s(in)}$ which are its components in the 
Laguerre-Gaussian basis.

In terms of the mode amplitudes, the dynamical equations 
for the injected OPO are \cite{fabre}
\begin{eqnarray}
&&\dfrac{\text{\textsl{d}} \alpha_{p}}{\text{\textsl{d}} t} = - ( \kappa_{p} + i \Delta_{p}) \alpha_{p} + \chi \left( \alpha_{+}^{s} \alpha_{-}^{i} + \alpha_{-}^{s} \alpha_{+}^{i} \right)  + \eta_{p} \alpha^{p}_{in} \nonumber\\
&&\dfrac{\text{\textsl{d}} \alpha^{s}_{+}}{\text{\textsl{d}} t} = - ( \kappa + i \Delta) \alpha_{+}^{s} - \chi \alpha_{-}^{i*} \alpha_{p} + \eta_{s} \alpha^{s(in)}_{+}   \nonumber\\
&&\dfrac{\text{\textsl{d}} \alpha^{s}_{-}}{\text{\textsl{d}} t} = - ( \kappa + i \Delta) \alpha_{-}^{s} - \chi \alpha_{+}^{i*} \alpha_{p} + \eta_{s} \alpha^{s(in)}_{-} \label{eqm}\\
&&\dfrac{\text{\textsl{d}} \alpha^{i}_{+}}{\text{\textsl{d}} t} = - ( \kappa + i \Delta) \alpha_{+}^{i} - \chi \alpha_{-}^{s*} \alpha_{p} \nonumber\\
&&\dfrac{\text{\textsl{d}} \alpha^{i}_{-}}{\text{\textsl{d}} t} = - ( \kappa + i \Delta) \alpha_{-}^{i} - \chi \alpha_{+}^{s*} \alpha_{p} \nonumber\quad,
\end{eqnarray} 
where $\kappa_{p}$ and $\Delta_{p}$ are, respectively, the cavity damping rate and the frequency detuning for the pump field; $\kappa$ and $\Delta$ 
are, respectively, the common cavity damping rate and frequency detunning for the down-converted fields; and the constants $\eta_{j}=\sqrt{T_{j}}/\tau_{j}$ $(j=p,s)$ are the coupling between the input fields and the fields inside the cavity. $T_{j}$ and $\tau_{j}$ are, respectively, the input mirror transmition coefficient and the cavity round trip time 
for each field. The nonlinear coupling constant $\chi$ rules the energy exchange between pump 
and down-converted fields. It is proportional to the spatial overlap 
between the transverse modes and to the effective 
second-order nonlinear susceptibility of the cristal.\cite{fabre}

\subsection{Free running OPO}
\label{freeopo}

It is interesting to consider first the case of a free running OPO, 
for which we take $\alpha_{\pm}^{s(in)}=0\,$. For simplicity, we shall assume resonant operation ($\Delta = \Delta_p = 0$). In this case, the steady state solution of the dynamical equations gives
\begin{eqnarray}
|\alpha^{s}_{+}|=|\alpha^{i}_{-}|=A \quad ,
\nonumber\\
|\alpha^{s}_{-}|=|\alpha^{i}_{+}|=B \quad ,
\end{eqnarray}
where $A$ and $B$ are unknown constants. Their individual values are not fixed by the dynamical equations and somehow depend on the initial conditions. However, the total signal and idler intensities are well defined, as we shall see shortly. 
This {\it degeneracy} is reminiscent of the implicit 
symmetry assumed in the dynamical equations. If cavity and/or 
crystal anisotropies are considered, then a rather more complicated scenario shows up as discussed in ref.\onlinecite{Martinelli}. 

Another degeneracy becomes evident when we consider the steady state values for the phases of the modes amplitudes. By taking $\theta^{p}_{in}=0\,$, we arrive at
\begin{eqnarray}
\theta^{p}=0 \quad ,
\nonumber\\
\theta^{s}_{+}+\theta^{i}_{-}=0 \quad ,
\\
\theta^{s}_{-}+\theta^{i}_{+}=0 \quad .
\nonumber\\
\end{eqnarray}
Again, the individual values of $\theta^{s}_{\pm}$ and $\theta^{i}_{\pm}$ are not fixed by the dynamical equations and should also depend on the initial conditions. 

\begin{figure}
\centering
\includegraphics[scale=0.8]{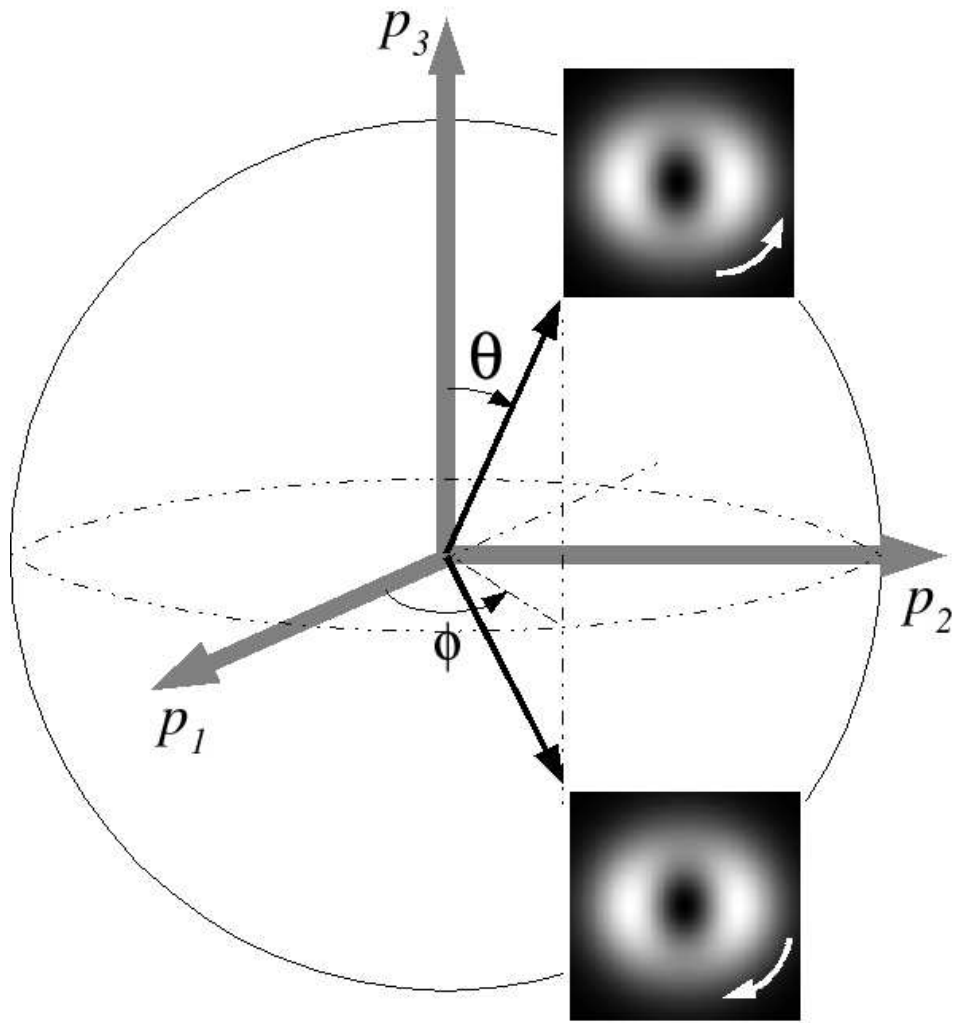}
\caption{Correlation imposed by optimal spatial overlap and orbital angular momentum conservation. Signal and idler are represented by two points on the Poincar\'e sphere symmetrically disposed with respect to the equatorial plane.}
\label{fig:2}
\end{figure}

It is instructive to look at these degeneracies in terms of the Poincar\'e representation of transverse modes. Solving the steady state equations for the intracavity pump intensity $I_p\equiv|\alpha_p|^2$ we obtain the usual clipping value
\begin{equation}
I_p = \left( \frac{\kappa}{\chi}\right)^2 \quad .
\end{equation}
The stationary value of the total intensity of each down converted field is
\begin{equation}
I_s=I_i=A^2+B^2=\frac{\chi}{\kappa_p}\left( \frac{\eta_p}{\kappa_p}
|\alpha^{p}_{in}| - \frac{\kappa}{\chi}\right) \quad ,
\end{equation}
where $I_j\equiv |\alpha^{j}_{+}|^2 + |\alpha^{j}_{-}|^2$ for 
$j=s,i\,$. Now, one can easily obtain the stationary values of the Stokes parameters for signal and idler:
\begin{eqnarray}
p^s_1=p^i_1&=&\frac{-2AB}{A^2+B^2}\cos\Delta\theta \quad ,
\nonumber\\
p^s_2=p^i_2&=&\frac{-2AB}{A^2+B^2}\sin\Delta\theta \quad ,
\label{correl}\\
p^s_3=-p^i_3&=&\frac{A^2-B^2}{A^2+B^2} \quad ,\nonumber
\end{eqnarray}
where 
$\Delta\theta\equiv \theta^s_+ - \theta^s_- = \theta^i_+ - \theta^i_-$. 
Therefore, the Stokes parameters of signal and idler fields correspond to points on the Poincar\'e sphere symmetrically disposed with respect to the equatorial plane, as described in fig.\ref{fig:2}. Since $A$, $B$ and $\Delta\theta$ are not fixed, they are free to diffuse and the steady state for the down converted fields can fall at any pair of points on the sphere, as far as they respect this symmetry. Physically, it means that signal and idler have the same intensity distribution, and therefore optimal spatial overlap, but opposite helicities due to OAM conservation. 

Of course, conservation laws are connected to symmetries. In the dynamical equations (\ref{eqm}) we have implicitly assumed that the OPO cavity and crystal present cylindrical symmetry. The main anisotropy present in an OPO is the crystal birrefringence. In ref.\onlinecite{Martinelli}, the operation of a type-II OPO driven by a Laguerre-Gaussian pump was observed for the first time with transfer of OAM to the idler beam, but not to the signal. For a type-II OPO, signal and idler have orthogonal polarizations, and astigmatic effects induced by the crystal birrefingence prevents the OAM transfer to the signal beam.
In fact, as discussed in ref.\onlinecite{Martinelli}, a birrefringence dependent astigmatism will cause a frequency split between transverse modes that conserve OAM. For this reason, we will propose an experimental setup based on a type-I OPO, where signal and idler have the same polarization.

The treatment of the nondeterministic dynamics with the addition of noise terms to the dynamical equations will be left for a future investigation. However, we can built a physical picture from the Poincar\'e representation which allows us to infere some of the main characteristics of the noisy evolution. In fact, the addition of noise should drive the down converted fields along random trajectories in the Poincar\'e sphere. This is analogous to the random trajectory delineated by the electric field phasor in a geometric representation of laser phase diffusion in the complex plane. In analogy, we can think of random trajectories in the Poincar\'e sphere as a {\it mode diffusion}. Of course, higher order modes may play an important role in the noisy evolution of the free running OPO, and a considerable improvement of the model might be necessary. However, in the case of an injected OPO, higher order modes may still contribute to noise but not to the macroscopic behavior of the amplified modes.

\subsection{Injected OPO}
\label{injopo}

Let us now consider a signal seed prepared in an arbitrary superposition of first order modes represented on the Poincar\'e sphere by the polar and azimuth angles $\theta$ and $\phi$, as shown in 
fig. \ref{fig:2}. The injected signal is then represented by the following amplitudes: 
\begin{eqnarray}
\alpha_{+}^{s(in)}&=&\sqrt{I_{s}^{in}} \cos \left( \frac{\theta}{2} \right) e^{-i \frac{\phi}{2}}\quad ,  \label{alphas+}\\
\alpha_{-}^{s(in)}&=&\sqrt{I_{s}^{in}} \sin\left( \frac{\theta}{2} \right) e^{i \frac{\phi}{2}} \quad .
\label{alphas-}
\end{eqnarray}

With this choice for the injection field, the dynamical equations 
are simplified by the definition of a new set of transformed amplitudes 
$\alpha_{j}$ and $\alpha^{\prime}_{j}$ $(j=s,i)$ given by:

\begin{eqnarray}
\nonumber \alpha_{s} &=& \cos\left( \frac{\theta}{2} \right) e^{i \frac{\phi}{2}}  \alpha^{s}_{+} +  \sin\left( \frac{\theta}{2} \right) e^{-i \frac{\phi}{2}} \alpha^{s}_{-}  \\
\nonumber \alpha_{i} &=& \sin\left( \frac{\theta}{2} \right) e^{i \frac{\phi}{2}}  \alpha^{i}_{+} +  \cos\left( \frac{\theta}{2} \right) e^{-i \frac{\phi}{2}} \alpha^{i}_{-} \\
\alpha^{\prime}_{s} &=& -\sin\left( \frac{\theta}{2} \right) e^{i \frac{\phi}{2}}  \alpha^{s}_{+} +  \cos\left( \frac{\theta}{2} \right) e^{-i \frac{\phi}{2}} \alpha^{s}_{-}  \label{change}\\
\nonumber
\alpha^{\prime}_{i} &=& \cos\left( \frac{\theta}{2} \right) e^{i \frac{\phi}{2}}  \alpha^{i}_{+} -  \sin\left( \frac{\theta}{2} \right) e^{-i \frac{\phi}{2}} \alpha^{i}_{-} \quad .
\end{eqnarray}

The dynamical equations can be rewritten in terms of the transformed amplitudes 
and one can easily show that the steady state solutions for $\alpha^{\prime}_{j}$ 
vanish (see appendix). 
Therefore, only three complex amplitudes $\alpha_{p}$, $\alpha_{s}$ 
and $\alpha_{i}$ are left.

The new equations of motion are
\begin{eqnarray}\label{alfasdin}
\nonumber && \dfrac{d \alpha_{p}}{dt} = - (\kappa_{p} + i \Delta_{p}) \alpha_{p} + \chi  \alpha_{s} \alpha_{i} + \eta_{p} \alpha^{in}_{p} \\
&& \dfrac{d \alpha_{s}}{dt} = - (\kappa + i \Delta) \alpha_{s} - \chi  \alpha^{*}_{i} \alpha_{p}  + \eta_{s} \sqrt{I_{s}^{in}} \\
\nonumber && \dfrac{d \alpha_{i}}{dt} = - (\kappa + i \Delta) \alpha_{i} - \chi  \alpha^{*}_{s} \alpha_{p}  \quad .
\end{eqnarray}

The stationary behavior is obtained from the steady-state 
solutions of eqs. (\ref{alfasdin}). We shall consider, for simplicity, 
that the interacting fields are resonant with the optical cavity 
($\Delta = \Delta_p = 0$). 
By setting the time derivatives equal to zero in the dynamical 
equations, we arrive at a set of algebraic equations which can be 
solved for the steady-state amplitudes. 

The steady-state solutions for the down-converted modes can be put 
in a simple form in terms of the pump mode steady-state solution,
\begin{eqnarray}
\alpha_{s} &=& \frac{\eta_{s} \rho \sqrt{I_{s}^{in}}}{(\kappa^{2} - \chi^{2}|\alpha_{p}|^{2})} \\
\alpha_{i} &=& \frac{-\eta_{s} \chi \sqrt{I_{s}^{in}} \alpha_{p}}
{(\kappa^{2} - \chi^{2}|\alpha_{p}|^{2})} \quad .
\end{eqnarray} 

The solutions for the modulus $|\alpha_{p}|$ of the pump mode amplitude are given by the roots of a fifth order polynomial
\begin{eqnarray}\label{fifth}
\nonumber & b^{2}|\alpha_{p}| + \left( |\alpha_{p}| - a\right)  \left( |\alpha_{p}| - \dfrac{\kappa}{\chi}\right) ^{2}\left( |\alpha_{p}| + \dfrac{\kappa}{\chi}\right) ^{2} = 0\quad ,
\end{eqnarray} 
where $a=\eta_{p} |\alpha_{p}^{in}|/\kappa_{p}$, and $b=\eta_{s} \kappa |\alpha_{s}^{in}|/(\chi \sqrt{\kappa \kappa_{p}})$.

There is no algorithm to find analytical expressions for the roots of a fifth order polynomial. However, simple approximate expressions can be given for some operation regimes of the OPO. In ref.\onlinecite{Coutinho} these expressions are presented together with a linear stability analysis showing that the stable solutions satisfy 
$|\alpha_{p}|<\kappa/\chi\,$.

\section{Geometric phase conjugation and an experimental proposal}
\label{geoconj}

The geometric phase was first discovered in 1956 by S. Pancharatnam.\cite{pancharatnam} This kind of geometric phase appears when a monochromatic light beam passes through a cyclic transformation on its polarization state, represented by a closed path on the Poincar\'e sphere. The geometric phase acquired by the beam is equal to minus one half the solid angle enclosed by the path.

In analogy to Pancharatnam phase, van Enk proposed the appearence of a geometric phase as we make cyclic transformations on transverse modes,\cite{Enk} even though the polarization state and the propagation direction of the beam do not change. Such transformations can be made with the help of astigmatic mode converters \cite{HELO} and spatial rotators (Dove prisms). 
In Galvez \textit{et al},\cite{Galvez} the existence of van Enk's geometric phase was first demonstrated through the interference between a first-order mode 
following cyclic mode conversions and a $TEM_{00}$ reference beam. As in the case of 
Pancharatnam's phase, the geometric phase aquired in the cyclic evolution is equal to minus 
one half the solid angle enclosed by the path followed in the Poincar\'e representation of 
first order spatial modes.

\begin{figure}
\centering
\includegraphics[scale=0.5]{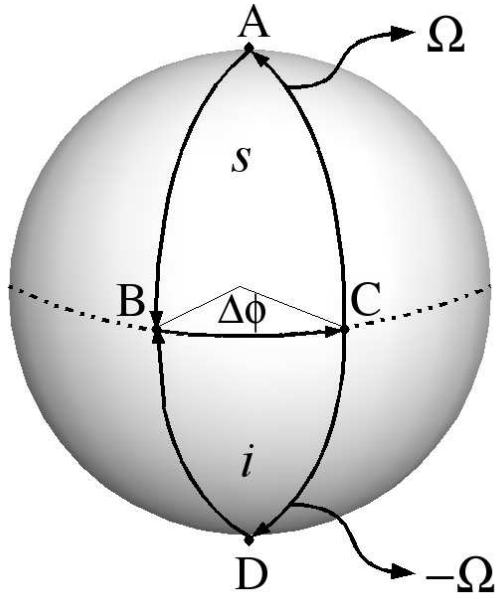}
\caption{Poincar\'e representation of the transformations performed on 
signal and idler beams.}
\label{fig:3}
\end{figure}

Now, let us consider the injected OPO described above. A cyclic transformation performed adiabatically on the injection mode, equivalent to a slow variation of $\theta$ and $\phi$ in eqs. (\ref{alphas+}) and (\ref{alphas-}), will cause the down converted beams to perform closed paths on the Poincar\'e sphere.
The Stokes parameters of the down-converted beams can be readily calculated by inverting eqs. (\ref{change}) and inserting the steady state solutions in definitions (\ref{p1})-(\ref{p3}):
\begin{eqnarray}
p^{s}_{1} &=& p^{i}_{1} = \sin \theta \cos \phi \quad ,\\
p^{s}_{2} &=& p^{i}_{2} = - \sin \theta \sin \phi \quad ,\\
p^{s}_{3} &=& - p^{i}_{3} = \cos \theta \quad .
\end{eqnarray}
Since signal and idler have opposite signs of $p_{3}$, they will perform symmetric paths with respect to the equator of the Poincar\'e sphere. Closed paths will be followed in opposite senses with respect to the unit vector normal to the sphere, as described in fig.\ref{fig:3}. Therefore, the geometric phase acquired by the idler field will be the conjugate of the one acquired by the signal. 

However, there is a subtle difference between this geometric phase and the one described in refs.\onlinecite{Enk,Galvez,poincaresphere}. In that case, the geometric phase is aquired along the propagation of the optical beam through a sequence of mode converters that implement a cyclic operation. Therefore, the time scale involved in the whole process is given by the time of flight of the optical beam through the setup and the notion of adiabatic transport is not clear. In our case, the idea is to drive the OPO adiabatically through a cyclic transformation during a time scale slow enough to consider the system approximately stationary at each moment. More precisely, we propose to inject a type-I OPO with an optical beam initially prepared in a given point of the Poincar\'e sphere, and to modify the setup adiabatically in order to drive the OPO slowly through a cyclic evolution. The relative geometric phase aquired by the down-converted beams can then be measured by their mutual interference. The choice of type-I phase match is important to avoid the astigmatic 
effects discussed above.

\begin{figure}
\centering
\includegraphics[scale=0.5]{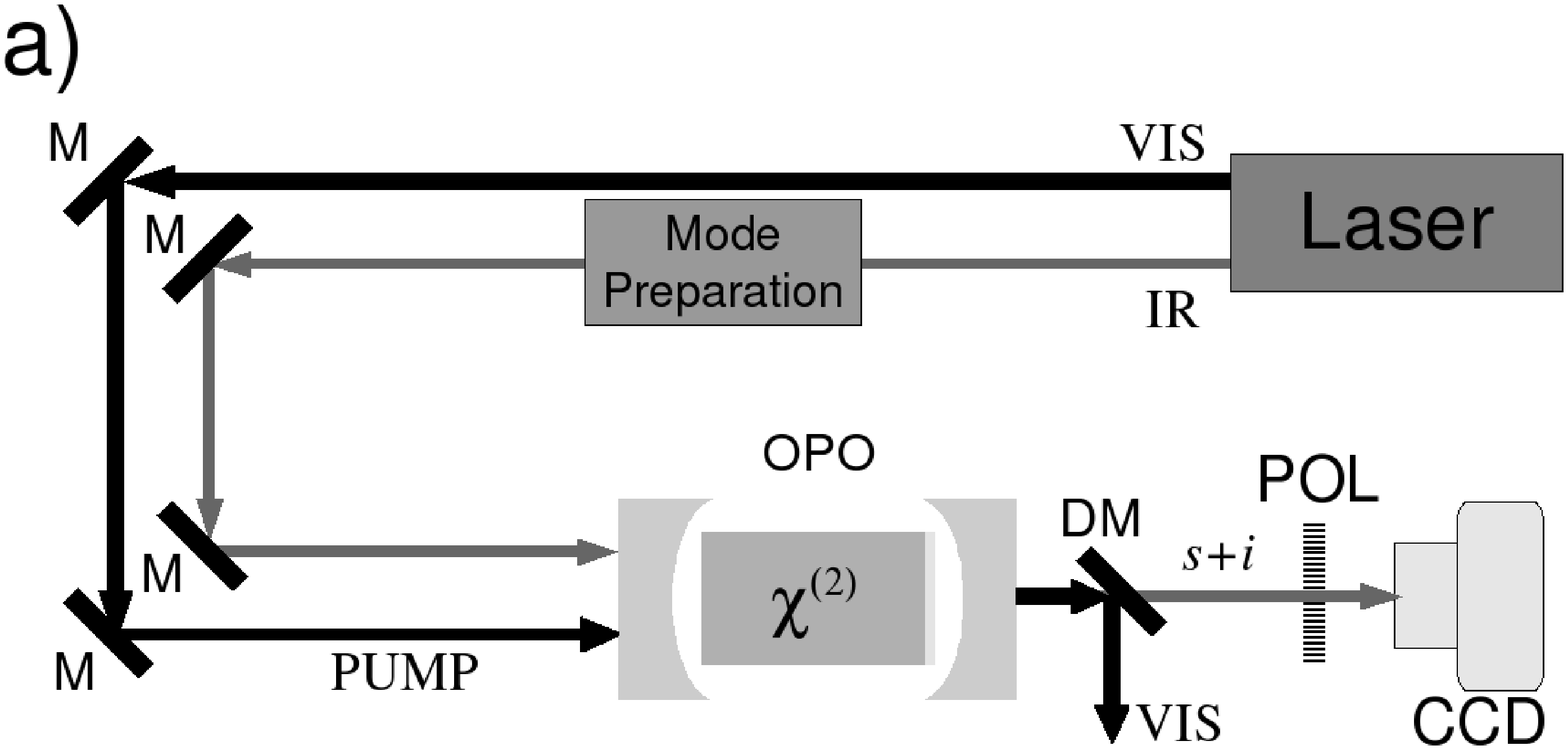}
\includegraphics[scale=0.5]{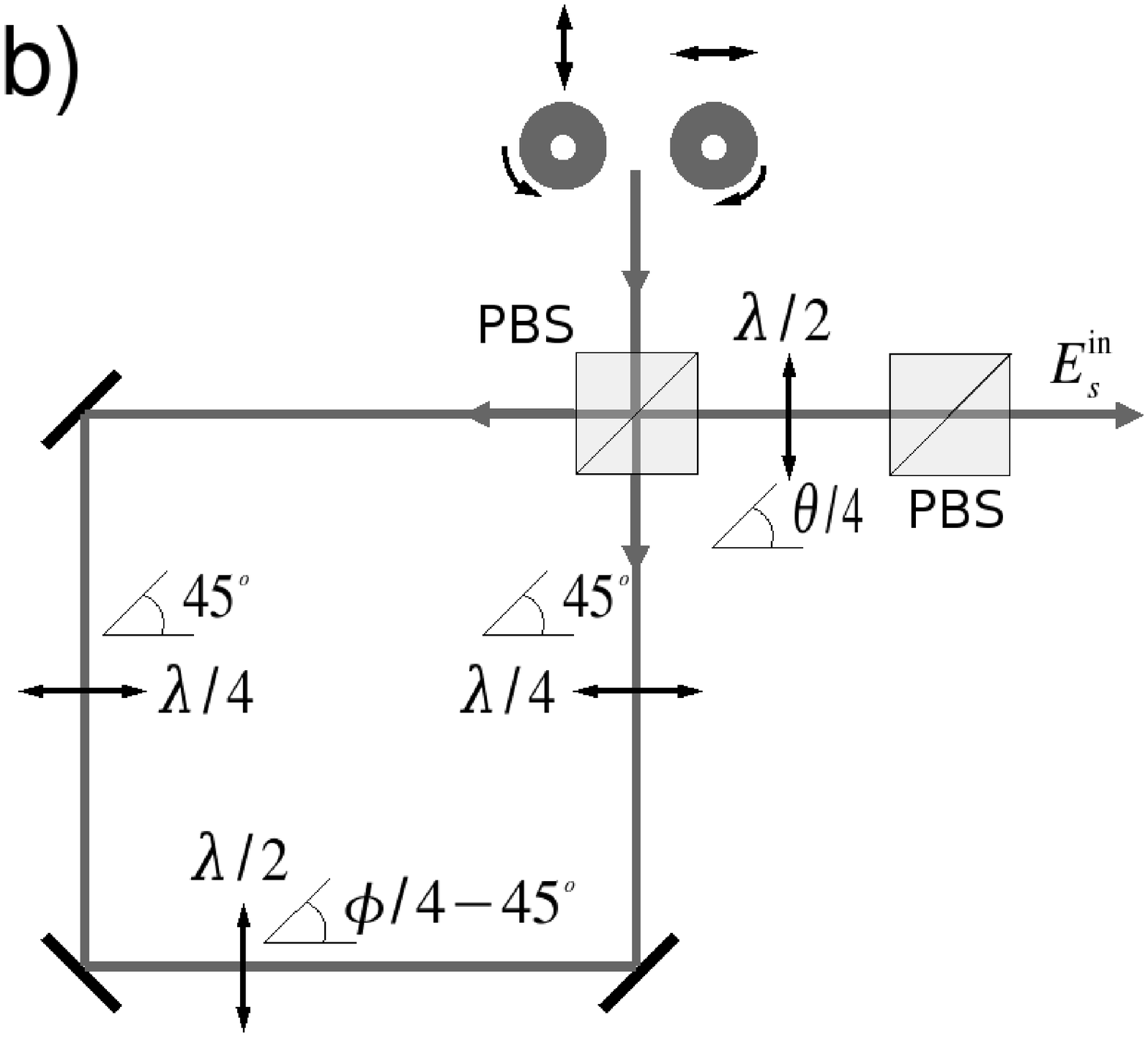}
\caption{a) A proposed experimental setup for measuring the geometric phase conjugation. 
M=mirror, DM=dichroic mirror, POL=polarizer. b) Mode preparation setup. PBS=polarizing 
beam splitter.}
\label{fig:4}
\end{figure}

Consider, for example, the setup sketched in fig.\ref{fig:4}a. A dual wavelength laser 
source provides both the OPO pump at the visible wavelength (VIS) and the injection 
seed at the infrared (IR). 
The injection beam is sent 
to a mode preparation setup where arbitrary modes on the Poincar\'e sphere can be produced. 
The mode preparation settings are then adiabatically varied in order to drive the OPO 
injection over a given path on the sphere. The adiabatic character of this evolution 
is important to ensure that the idler beam produced by the OPO preserves its symmetry 
with respect to the signal beam. Therefore, the mode preparation settings must be 
varied in a time scale much longer than the cavity lifetime $2\pi/\kappa$ for the down-converted beams. After the OPO, the IR beams are transmitted through a dichroic 
mirror (DM), where they are separated from the visible pump beam, and spatially 
interfere on the screen of a CCD camera. In the case of a type II OPO, the 
down-converted beams have orthogonal polarizations, so that a polarizer (POL) 
oriented at $45^o$ with respect to their polarizations must be introduced in order 
to provide interference. 

The mode preparation setup is described in fig.\ref{fig:4}b. Two Laguerre-Gaussian 
modes with opposite helicities and orthogonal polarizations are sent to a phase shifter 
which consists of a Sagnac interferometer with a polarizing beam splitter (PBS)
as the input/output port, two quarter-wave plates and one half-wave plate. 
The quarter-wave plates are oriented at $45^{o}$ with 
respect to the horizontal axis. The orientation of the half-wave plate can be varied 
and it is convenient to write it as $\phi/4 - 45^{0}\,$. With this configuration, 
the relative phase between the counter-propagating beams in the output of the phase 
shifter is $\phi\,$. Finally, after recombining at the output of the Sagnac 
interferometer, the two phase shifted modes pass through a second half-wave plate 
at a variable orientation $\theta/4$ and another polarizing beam splitter. The 
resulting field profile is
\begin{equation}
E_{s}^{in} = \sqrt{I_{s}^{in}}\left[
e^{-i\phi/2}\,\cos\left(\theta/2\right)\,\psi_{+}(\vec{\rho},z)+
e^{i\phi/2}\,\sin\left(\theta/2\right)\,\psi_{-}(\vec{\rho},z) 
\right]\;,
\end{equation}
which is equivalent to the injection amplitudes given by eqs.(\ref{alphas+}) 
and (\ref{alphas-}). Therefore, the injection field can be prepared at any point 
on the Poincar\'e sphere by proper orientation of the half wave plates. 

\begin{figure}
\centering
\includegraphics[scale=0.5]{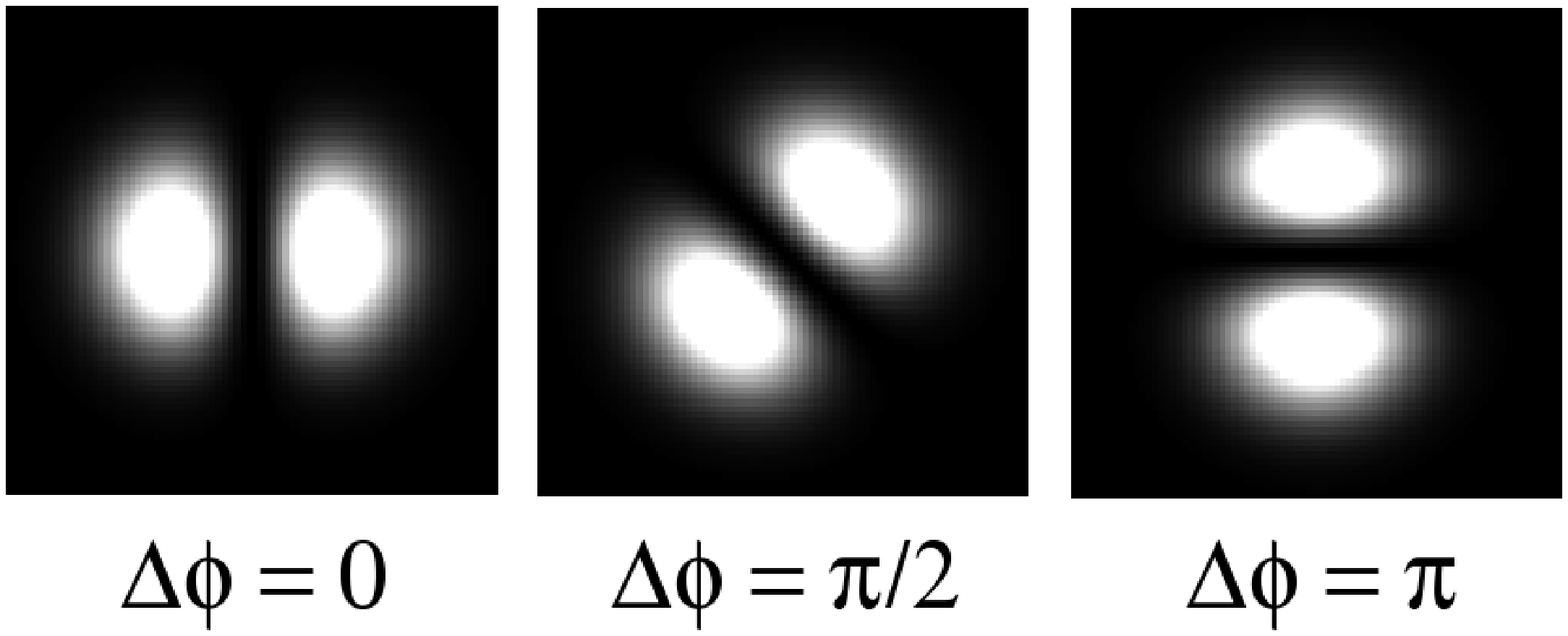}
\caption{Calculated intereference patterns for different cyclic 
evolutions of the OPO.}
\label{fig:5}
\end{figure}

After the cyclic transformation, the injection seed follows the path $ABCA$ 
(fig.\ref{fig:3}) corresponding to a solid angle $\Omega=\Delta\phi$ in the 
Poincar\'e sphere, 
where $\Delta\phi$ is the azimuthal angle enclosed by the path on the equator. 
The geometric phase aquired by the injected signal is $\gamma_s=-\Delta\phi/2\,$.
The idler beam is then 
adiabatically driven over the symmetric path $DBCD\,$, so that a 
geometric phase $\gamma_i=-\gamma_s=\Delta\phi/2$ is aquired by the idler. 
Therefore, the phase difference between the 
down-converted beams is increased by $\Delta\phi$ and can be measured 
by the change in their mutual interference pattern, 
as shown in fig.\ref{fig:5}. After the cyclic transformation, the memory of 
the adiabatic evolution is registered as a rotation of the mutual interference 
pattern. 

\section{Conclusions}
\label{final}

In conclusion, we studied the classical dynamics of an optical parametric oscillator injected with a signal seed initially prepared in a transverse mode carrying orbital angular momentum. By the use of a Poincar\'e sphere representation, we stablished a transverse mode symmetry between the down converted beams. We also predicted an effect of geometric phase conjugation when adiabatic mode conversions, following closed paths in the Poincar\'e sphere, are performed on the injected beam. An experimental setup for measuring this effect was proposed and one of its interesting aspects is the memory of the adiabatic evolution registered in the image of an interference pattern. The investigation of this effect in the quantum domain is left for a future work.

\acknowledgments

The authors acknowledge financial support from the Brazilian funding agencies CNPq 
(Conselho Nacional de Desenvolvimento Cient\'{\i}fico e Tecnol\'ogico), 
CAPES (Coordenadoria de Aperfei\c coamento de Pessoal de N\'{\i}vel Superior), and 
FAPERJ (Funda\c c\~ao de Amparo \`a Pesquisa do Estado do Rio de Janeiro). 
We also would like to thank P. H. Souto Ribeiro for fruitful discussions.

\section*{Appendix}

The transformed variables introduced by eqs. \ref{change} allow us to rewrite the 
dynamical equations with a single injection term:
\begin{eqnarray}
&& \dfrac{d \alpha_{p}}{dt} = - (\kappa_{p} + i \Delta_{p}) \alpha_{p} + \chi^{*}_{ef} \left( \alpha_{s} \alpha_{i} - \alpha^{\prime}_{s} \alpha^{\prime}_{i} \right)  + \eta_{p} \alpha^{in}_{p} \\
&& \dfrac{d \alpha_{s}}{dt} = - (\kappa + i \Delta) \alpha_{s} - \chi_{ef}  \alpha^{*}_{i} \alpha_{p}  + \eta_{s} \sqrt{I^{in}_{s}} \\
&& \dfrac{d \alpha_{i}}{dt} = - (\kappa + i \Delta) \alpha_{i} - \chi_{ef}  \alpha^{*}_{s} \alpha_{p}  \\
&& \dfrac{d \alpha^{\prime}_{s}}{dt} = - (\kappa + i \Delta) \alpha^{\prime}_{s} - \chi_{ef}  \alpha^{\prime *}_{i} \alpha_{p} \\
&& \dfrac{d \alpha^{\prime}_{i}}{dt} = - (\kappa + i \Delta) \alpha^{\prime}_{i} - \chi_{ef}  \alpha^{\prime *}_{s} \alpha_{p}  \quad .
\end{eqnarray}

This substantially simplifies the steady state solution, which can be readily found 
from the following algebraic system:
\begin{eqnarray}
&&  \alpha_{p} = \dfrac{\chi}{\kappa_{p}}  \left( \alpha_{s} \alpha_{i} - \alpha^{\prime}_{s} \alpha^{\prime}_{i} \right)  + \dfrac{\eta_{p}}{\kappa_{p}} \alpha^{in}_{p} \\
&& \alpha_{s} = - \dfrac{\chi}{\kappa} \alpha^{*}_{i} \alpha_{p} + \dfrac{ \eta_{s}}{\kappa} \sqrt{I^{in}_{s}} \\
&& \alpha_{i} = - \dfrac{\chi}{\kappa} \alpha^{*}_{s} \alpha_{p} \\
\label{nesta} && \alpha^{\prime}_{s} = - \dfrac{\chi}{\kappa} \alpha^{\prime *}_{i} \alpha_{p} \\
\label{esta} && \alpha^{\prime}_{i} = - \dfrac{\chi}{\kappa} \alpha^{\prime *}_{s} \alpha_{p} \quad ,
\end{eqnarray} 
where we assumed $\Delta_p = \Delta = 0$ for simplicity.

A straightforward algebra with the last two equations, gives
\begin{equation}
\alpha^{\prime}_{s} \left( 1 - 
\dfrac{|\chi|^{2}}{\kappa^{2}} |\alpha_{p}|^{2} \right) = 0 \quad .
\end{equation}
One of the solutions is $|\alpha_{p}|^{2}=\kappa^{2}/\chi^{2}$, which is nonphysical 
since it corresponds to the noninjected case solution for the pump mode. The other one 
is  $\alpha^{\prime}_{s}=0$, which shows that both amplitudes $\alpha^{\prime}_{s}$ 
and  $\alpha^{\prime}_{i}$ vanish. Therefore, the corresponding modes are not amplified 
and do not oscillate. However, they will certainly contribute to noise but this 
will be left for a future investigation.


\newpage

\begin{thebibliography}{99}

\bibitem{arlt}
J. Arlt, K. Dholakia, L. Allen, and M. J. Padgett,
"Parametric down conversion for light beams possessing orbital angular momentum,"
Phys. Rev. A \textbf{59}, 3950-3952 (1999).

\bibitem{Mair} 
A. Mair, A. Vaziri, G. Weihs, and A. Zeilinger,
"Entanglement of the orbital angular momentum states of photons," 
\textit{Nature} (London) {\textbf 412}, 313-316 (2001).

\bibitem{Caetano} D.P. Caetano, M.P. Almeida, P.H. Souto Ribeiro, 
J.A.O. Huguenin, B. Coutinho dos Santos, and A.Z. Khoury, 
"Conservation of orbital angular momentum in stimulated down-conversion,"
Physical Review A {\textbf 66}, 041801 (Rapid Comm.) (2002).

\bibitem{nossoprl} 
P. H. Souto Ribeiro, D. P. Caetano, M. P. Almeida, J. A. Huguenin, 
B. Coutinho dos Santos, and A. Z. Khoury,
"Observation of image transfer and phase conjugation in stimulated down-conversion,"
\textit{Phys. Rev. Lett.} \textbf{87}, 133602 (2001).

\bibitem{Martinelli} 
M. Martinelli, J. A. O. Huguenin, P. Nussenzveig, and A. Z. Khoury, 
"Orbital angular momentum exchange in an optical parametric oscillator,"
\textit{Phys. Rev. A} \textbf{70}, 013812 (2004).

\bibitem{berry} 
M. V. Berry, 
"Quantal phase-factors accompanying adiabatic changes,"
{\textit Proc. R. Soc. London A} \textbf{392}, 45 (1984).

\bibitem{HELO} 
M. W. Beijersbergen, L. Allen, H. E. L. O. var der Veen, and J. P. Woerdman,
"Astigmatic laser mode converters and transfer of orbital angular momentum,"
\textit{Opt. Commun.} \textbf{96}, 123-132 (1993).

\bibitem{russos}
E. Abramochkin and V. Volostnikov, 
"Beam transformations and nontransformed beams,"
\textit{Opt. Commun.} {\bf 83}, 123-135 (1991).

\bibitem{Enk} 
S. J. van Enk, 
"Geometric phase, transformations of Gaussian light-beams and angular-momentum transfer,"
\textit{Opt. Commun} \textbf{102}, 59 (1993).

\bibitem{Galvez} 
E. J. Galvez,  P. R. Crawford, H. I. Sztul, M. J. Pysher, P. J. Haglin, and R. E. Williams, 
"Geometric phase associated with mode transformations of optical beams bearing orbital angular momentum,"
\textit{Phys. Rev. Lett.} \textbf{90}, 203901 (2003).

\bibitem{poincaresphere}
M. J. Padgett, and J. Courtial, 
"Poincare-sphere equivalent for light beams containing orbital angular momentum,"
Optics Letters {\textbf 24}, 430, (1999).

\bibitem{pancharatnam} 
S.Pancharatnam, 
"Generalized theory of interference and its applications. Part I. Coherent pencils,"
\textit{Proc. Ind. Acad. Sci.} \textbf{44}, 247 (1956).

\bibitem{quantcomp1}
J. A. Jones, V. Vedral, A. Ekert, and G. Castagnoli,
"Geometric quantum computation using nuclear magnetic resonance," 
\textit{Nature} (London) {\textbf 403}, 869-871 (2000).

\bibitem{quantcomp2}
L.-M. Duan, J. I. Cirac, and P. Zoller,
"Geometric manipulation of trapped ions for quantum computation," 
\textit{Science} {\textbf 292}, 1695-1697 (2001).

\bibitem{brendel}
J. Brendel, W. Dultz, and W. Martienssen, 
"Geometric phases in 2-photon interference experiments,"
\textit{Phys. Rev. A} \textbf{52}, 2551-2556 (1995).

\bibitem{vaziri}
A. Vaziri, G. Weihs, and A. Zeilinger,
"Experimental two-photon, three-dimensional entanglement for quantum communication,"
\textit{Phys. Rev. Lett.} \textbf{89}, 240401 (2002).

\bibitem{steve}
S. P. Walborn, A. N. de Oliveira, R. S. Thebaldi, and C. H. Monken, 
"Entanglement and conservation of orbital angular momentum in spontaneous parametric down-conversion,"
\textit{Phys. Rev. A} \textbf{69}, 023811 (2004).

\bibitem{Barbosa} 
H. H. Arnaut and G. A. Barbosa, 
"Orbital and intrinsic angular momentum of single photons and entangled pairs of photons generated by parametric down-conversion,"
\textit{Phys. Rev. Lett.} \textbf{85}, 286 (2000).

\bibitem{silvania}
Z. Y. Ou, S. F. Pereira, and H. J. Kimble,
"Realization of the Einstein-Podolsky-Rosen paradox for continuous-variables 
in nondegenerate parametric amplification,"
\textit{App. Phys. B} \textbf{55}, 265-278 (1992);
Z. Y. Ou, S. F. Pereira, J. Kimble, and K. C. Peng, 
"Realization of the Einstein-Podolsky-Rosen paradox for continuous-variables,"
\textit{Phys. Rev. Lett.} \textbf{68}, 3663-3666 (1992).

\bibitem{bowen}
W. P. Bowen, N. Treps, R. Schnabel, and P. K. Lam, 
"Experimental demonstration of continuous variable polarization entanglement,"
\textit{Phys. Rev. Lett.} \textbf{89}, 253601 (2002);
W. P. Bowen, R. Schnabel, P. K. Lam, and T. C. Ralph, 
"Experimental investigation of criteria for continuous variable entanglement,"
\textit{Phys. Rev. Lett.} \textbf{90}, 043601 (2003).

\bibitem{chiao}
W. R. Tompkin, M. S. Malcuit, R. W. Boyd, and R. Y. Chiao, 
"Time reversal of Berry's phase by optical phase conjugation,"
\textit{J. Opt. Soc. Am. B} \textbf{7}, 230-233 (1990).

\bibitem{fabre} 
C. Schwob, P. F. Cohadon, C. Fabre, M. A. M. Marte, H. Ritsch, A. Gatti, and L. Lugiato, 
"Transverse effects and mode couplings in OPOS,"
\textit{Appl. Phys. B} \textbf{66}, 685-699 (1998).

\bibitem{Coutinho} 
B. Coutinho dos Santos, K. Dechoum, A. Z. Khoury, L. F. da Silva, and
M. K. Olsen,
"Quantum analysis of the nondegenerate optical parametric oscillator with injected signal,"
\textit{Phys. Rev. A} \textbf{72}, 033820 (2005).

\end{thebibliography}
\end{document}